\documentclass[12pt]{aipproc} 
\layoutstyle{6x9} 
\usepackage{graphicx}
\usepackage{epsf}
\usepackage{epsfig}
\usepackage{color}

\newcommand{\be}{\begin{equation}}
\newcommand{\ee}{\end{equation}}
\newcommand{\bea}{\begin{eqnarray}}
\newcommand{\eea}{\end{eqnarray}}

\let \to=\rightarrow

\newcommand{\piA}{\ensuremath{\pip_{\sst A}}}
\newcommand{\piB}{\ensuremath{\pip_{\sst B}}}
\newcommand{\SA}{\ensuremath{s}}

\newcommand{\SAk}{\ensuremath{s_k}}

\newcommand{\swave}{\ensuremath{ S\hbox{-wave}}}
\newcommand{\pwave}{\ensuremath{ P\hbox{-wave}}}
\newcommand{\dwave}{\ensuremath{ D\hbox{-wave}}}
\newcommand{\sdash}{\ensuremath{ S\hbox{-}}}
\newcommand{\pdash}{\ensuremath{ P\hbox{-}}}
\newcommand{\ddash}{\ensuremath{ D\hbox{-}}}
\newcommand{\KAs}{\ensuremath{\kappa}}
\newcommand{\KBs}{\ensuremath{K_0^*(1430)}}
\newcommand{\KAp}{\ensuremath{K^*(892)}}
\newcommand{\KBp}{\ensuremath{K_1^*(1410)}}
\newcommand{\KCp}{\ensuremath{K_1^*(1680)}}
\newcommand{\KAd}{\ensuremath{K_2^*(1430)}}

\newcommand{\gammaz}{\ensuremath{\gamma_{0}}}

\newcommand{\sst}{\scriptscriptstyle}

\newcommand{\half}{\ensuremath{{1\over 2}}}

\newcommand{\MeVcc}{\ensuremath{\hbox{MeV}/c^2}}
\newcommand{\GeVcc}{\ensuremath{\hbox{GeV}/c^2}}

\let\mathrm=\rm

\newcommand{\pip}{\pi^+}

\newcommand{\Km}{K^-}

\newcommand{\Dp}{D^+}

\begin{document}

 \title{Measurement of the $\Km\pip$ \swave\ System in 
        $\Dp\to\Km\pip\pip$ Decays from Fermilab E791} 
 \author{B. Meadows \\
 {\small\sl University of Cincinnati, Cincinnati, OH, 45221, USA}}
 {address={Representing the Fermilab E791 Collaboration}}
 \begin{abstract}
  A new approach to the analysis of three body decays is presented.
  Measurements of the \swave\ $K\pi$ amplitude are made in independent
  ranges of invariant mass from threshold up to the upper kinematic
  limit in $\Dp\to\Km\pip\pip$ decays.  These are compared with 
  results obtained from a fit where
  the \swave\ is assumed to have \KAs\ and \KBs\ resonances.  Results
  are also compared with measurements of $\Km\pip$ elastic scattering.
  Contributions from $I=\half$ and $I={3\over 2}$ 
  are not resolved in this study.  If $I=\half$ dominates, however, the 
  Watson theorem prediction, that the phase behavour below
  $K\eta^{\prime}$ threshold should match that in elastic scattering,
  is not well supported by these data.  Production of $\Km\pip$ from
  these $D$ decays is also studied.
 \end{abstract}
 \footnote{Preprint numbers: FERMILAB-CONF-05-467-E, UCHEP-05-06}
 \maketitle

\section{\bf Introduction}
Decays of $D$ and $B$ mesons show promise as a source of 
information on the light-quark mesons they produce.  Their decays to
\swave\ systems
in three pseudo-scalar final states, may help to
improve our knowledge of the particularly confusing scalar meson
($J^P=0^+$) spectrum.  Until now, extracting such information has
been done in model-dependent ways that make assumptions about the
scalar states observed that can influence the results.  With large,
clean samples of such decays, anticipated to be coming from the
$B$ factories and the Tevatron collider, the need for new
approaches is a priority.

Knowledge of strange scalar mesons has relied on measurements of
\swave\ $\Km\pip$ scattering.  These come principally from SLAC
experiment E135 (LASS)
\cite{Aston:1987ir}
and cover the invariant mass range above 825~\MeVcc.  Data from
other experiments below this range exist, but with less precision
\cite{Estabrooks:1977xe}.
More measurements in the low mass region are required if the
possibility of the existence of a $\kappa$ state is to be
properly evaluated.

In this paper, a study of the decays
\cite{conjugate}
$\Dp\to\Km\pip\pip$ observed in data from Fermilab experiment E791
is presented.  In an earlier analysis
\cite{Aitala:2002kr}
the $\Km\pip$ \swave\ was modelled with Breit-Wigner (BW)
amplitudes for \KAs\ and \KBs\ resonances.
A less model-dependent analysis is presented here.  \swave\ amplitudes
are measured without assuming any specific dependence on $\Km\pip$
invariant mass, or on the presence of scalar resonances.  Results
are compared with the amplitude from our earlier analysis, and
also with measurements from LASS.

\section{Data Sample}
\begin{figure}[hbt]
 \centerline{%
 \epsfig{file=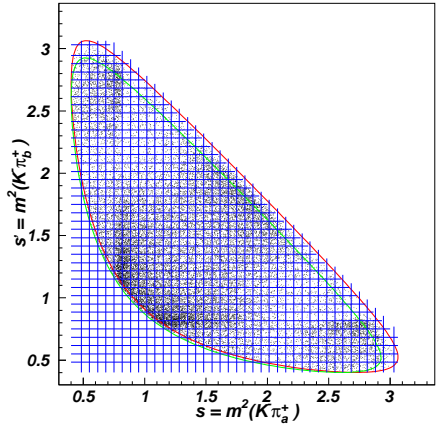,width=0.43\textwidth,angle=0}}
 \caption{Dalitz plot for $\Dp \to \Km\piA\piB$ decays.  The
  squared invariant mass of one $\Km\pip$ combination is
  plotted against the other
  combination.  The plot is symmetrized, each event
  appearing twice.  Lines in both directions indicate values, equally
  spaced in squared effective mass, at each of which the \swave\
  amplitude is determined by the method described in the text.
  Kinematic boundaries for the Dalitz plot are
  drawn for three-body mass values $M=1.810$ and $M=1.890$~\GeVcc,
  between which data are selected.
  \label{fig:dalitz_plot}}
\end{figure}

The selection process for events used in this paper is described
in Ref.~\cite{Aitala:2002kr}.  A signal consisting of 15,079 
$\Dp\to\Km\pip_a\pip_b$ decays, with a purity of $\sim 94$\%, is
obtained.  Fig.~\ref{fig:dalitz_plot} shows the Dalitz plot with
the $\Km\pip_a$ invariant mass squared plotted vs. that for the
$\Km\pip_b$ system.
Horizontal (and symmetrized vertical) bands corresponding
to the \KAp\ resonance are clearly seen.  A complex pattern
of both constructive and destructive interference
is seen near 2 (\GeVcc)$^2$ due to presence of \KBs, \KBp\,
and \KAd.
Evidence for \KCp, difficult to see due to smearing of the Dalitz
plot boundary resulting from the finite resolution in the
three-body $\Dp$ mass, may also exist.

An asymmetry along the \KAp\ bands, most easily described by
interference with a significant \swave\ component to these
decays, is also observed.  Information on this \swave\ amplitude
is obtained from its interference with the \KAp, and also the
other well-established resonances in the Dalitz plot.

\section{Method}\label{sec:method}
In Ref.~\cite{Aitala:2002kr}, as in most earlier analyses of
$D$ decays to three pseudo-scalar particles $ijk$, the
isobar model, with BW resonance forms, is used.
The decay amplitude ${\cal A}$ is described by
a sum of quasi two-body terms $D\to R+k,~R\to i+j$, in each of
the three channels $k=1,2,3$:
 \bea\label{eq:isobarmodel}
   {\cal A} &=& d_0 e^{i\delta_0} + 
        \sum_{n=1}^N d_{n} e^{i\delta_{n}}
        ~{F_{\sst R}(p,r_{\sst R},J)\over
        m_{\sst R_n}^2-\SA-im_{\sst R_n}\Gamma_{\sst R_n}(\SA)}
        \times
        F_{\sst D}(q,r_{\sst D},J)
        ~M_J(p,q)
 \eea
The squared invariant mass of the $ij$ system is $\SA$,
$J$ is the spin, $m_{\sst R_n}$ the mass and $\Gamma_{\sst R_n}(\SA)$
the width of each of the $N$ resonances $R_n$ seen to be
contributing to the decay.  For $J>0$, $F_{\sst R}$ and $F_{\sst D}$
are Blatt-Weisskopf form factors
\cite{BlattWeiss},
with effective radius parameters
$r_{\sst R}$ and $r_{\sst D}$, for all $R_n$ and for the parent $D$
meson, respectively.  For $J=0$, a Gaussian form suggested by Tornqvist
\cite{Tornqvist:1995kr}
is used for the $D$.  The momenta, $\vec p$ and $\vec q$ for $i$ 
and $k$, respectively, are defined in the $ij$ rest frame, and
$M_J(p,q) = (-2pq)P_J(\hat p\cdot\hat q)$ is introduced to
describe spin conservation in the decay.  The complex
coefficients $d_n e^{i\delta_n}~(n=0,N)$ depend on the $D$ decay and
are determined by a fit to the data.
The first term describes non-resonant ($NR$) decay to ${i+j+k}$
with no
intermediate resonance, and is assumed to be independent of $\SA$.
For $\Dp\to\Km\pip_a\pip_b$ decays we Bose-symmetrize
${\cal A}$ with respect to interchange of $\pip_a$ and $\pip_b$.

In Ref.~\cite{Aitala:2002kr}, an excellent fit to the data is
obtained with \KAs\ and \KBs\ resonant terms and the $NR$ term
comprising the $J=0$ part of Eq.~(\ref{eq:isobarmodel}).

In this paper, the
$\Km\pip$ \swave\ is examined in a less model-dependent way,
outlined here.  A more detailed account is given in
\cite{Aitala:2005yh}.

Terms appearing in Eq.~(\ref{eq:isobarmodel}) are grouped
according to the value of $J$.
The \swave\ part (all terms with $J=0$, including the $NR$
term) is factored
\bea\label{eq:swave}
   {\cal S} &=& \hbox{S}(\SA)
        \times
        M_0(p,q)
        F_{\sst D}(q,r_{\sst D},0)
\eea
into a partial wave $\hbox{S}(\SA)$, describing $\Km\pip$ scattering,
and the product $M_0(p,q)F_{\sst D}(q,r_{\sst D})$
describing the $D$ decay.
The \pdash\ and \ddash\ ($J=1,2$, respectively) waves are
factored in the same way:
\bea\label{eq:refwaves}
   {\cal P} &=& \hbox{P}(\SA)
        \times
        M_1(p,q)
        F_{\sst D}(q,r_{\sst D},1)
   ~;~
   {\cal D} ~=~ \hbox{D}(\SA)
        \times
        M_2(p,q)
        F_{\sst D}(q,r_{\sst D},1),
\eea
with partial waves $\hbox{P}(\SA)$ and $\hbox{D}(\SA)$
consisting of resonant terms given as in
Eq.~(\ref{eq:isobarmodel}).  The \swave\ $\hbox{S}(\SA)$,
however, is replaced by a set of values
$c_k e^{i\gamma_k}$ defined at 40 discrete squared invariant masses
$\SA=\SAk$.  These are indicated by the
lines in Fig.~\ref{fig:dalitz_plot}.  
The $c_k$ and $\gamma_k$ values are regarded as independent
parameters to be determined by the data.

An unbinned likelihood fit is made to the data, in a similar
way to that described in
Ref.~\cite{Aitala:2002kr}.
An incoherent function describing the 6\% background in the
sample from events that are not true $D$ decays is added in
proper proportion at each three-body $\Km\pip\pip$ mass to
a signal distribution proportional to $|{\cal A}|^2$.  There
are 86 parameters - all $(c_k,\gamma_k)$ and the coefficients
$d_k e^{i\delta_k}$ for \KAp, \KCp\ in the \pwave\ and \KAd\
in the \dwave.  For the \KAp, $d_k e^{i\delta_k}=1$ is used 
to provide the reference amplitude.

\section{Results and comparison with model-dependent fit}
\begin{figure*}[hbt]
 \centerline{%
 \epsfig{file=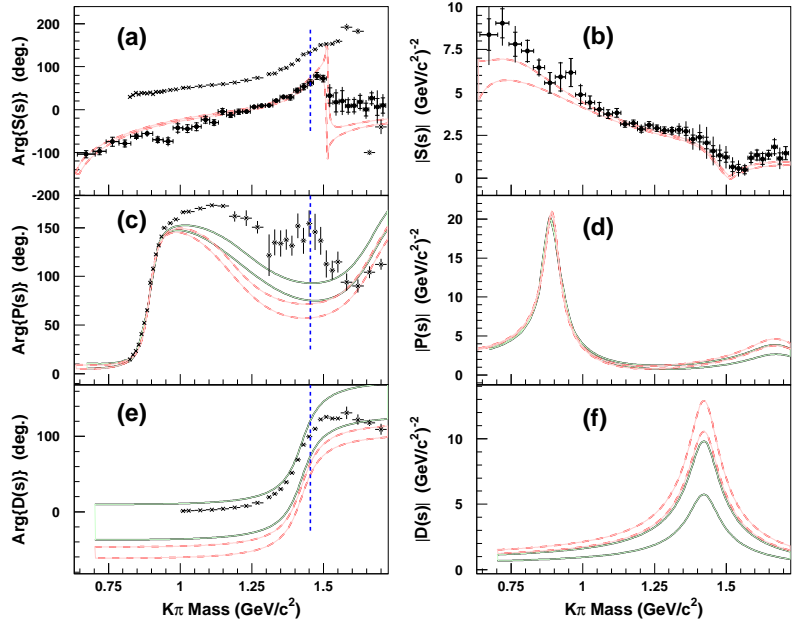,
         height=0.430\textheight,angle=0}}
 \caption{(a) Phases $\gamma_k=\arg{S(\SAk)}$ and (b) magnitudes
  $c_k=|S(\SAk)|$ of \swave\ amplitudes for $\Km\pip$ systems from
  $\Dp\to\Km\pip\pip$ decays with the amplitude and phase of the \KAp\ 
  as reference.  Solid circles, with error bars, show the values
  obtained from the model-independent fit described in the text.  
  The effect of adding systematic uncertainties in quadrature is
  indicated by extensions on the error bars.
  The magnitudes plotted include a $\Dp$ form-factor 
  $F_D(q,r_D,0)=e^{-25q^2/12}$.
%
  The phase and magnitude of P(\SA) are shown in (c) and (d), 
  respectively.  In each case, the parameters and error matrix 
  from the fit are used in Eq.~(\ref{eq:refwaves}) to produce 
  the solid curves shown one standard deviation above and
  below the central values.
%
  Similar plots for the
  \dwave\ amplitude D(\SA) are shown in (e) and (f).
  In all plots, the dashed curves show similar, one standard
  deviation limits for the amplitudes obtained from the isobar model
  fit in Ref.~\cite{Aitala:2002kr}.
  In (a), (c) and (e), $I=\half$ phases for $\Km\pip$ scattering
  measured in the LASS experiment
  are shown as
  $\times$'s with error bars indicating statistical uncertainties.
 \label{fig:solution0}}
\end{figure*}

This fit also results in an excellent description of the data.
Comparison of the observed and predicted population of the
Dalitz plot gives a $\chi^2$ probability of 50\% for 363 bins.
The \sdash, \pdash\ and \dwave s resulting from the fit are shown
in Fig.~\ref{fig:solution0}.  

These results are compared with the model-dependent fit
from Ref.~
\cite{Aitala:2002kr}
which provided an excellent description of the observed
Dalitz plot distribution.  Partial waves for this fit are
also shown in Fig.~\ref{fig:solution0}.
The main \swave\ features of both fits agree well.  Some 
differences, particularly at the highest and lowest ends of the
invariant mass range, result from shifts in the \pdash\ and
\dwave s.  Resonant
fractions and the total \swave\ fraction (about 75\%) also
agree well within statistical limits.

It can be concluded that, with the present sample size, no
significant distinction between the model-dependent and
model-independent parametrizations of the \swave\ can be
made.

A comparison of the \swave\ amplitudes $\hbox{S}(\SA)$
measured here with the amplitudes $T(\SA)$ obtained in $\Km\pip$
scattering is now made.
For each partial wave $J$ (for each iso-spin $I$)
it is expected that 
$\hbox{S}(\SA)=Q(\SA)\hbox{T}(\SA)/F_{\sst D}(q,r_{\sst D},J))
{\sqrt{\SA}/ p^{(J+1)}}$
where $Q(\SA)$ describes the \SA-dependence of $\Km\pip$
production in $D$ decays.  The Watson theorem
\cite{Watson:1952ji}
requires, provided there is no re-scattering of the
$\Km\pip_a$ from $\pip_b$, that $Q(\SA)$ is a real function, so
that phases found in $D$ decay should match those in $\Km\pip$
elastic scattering data.

$I=1/2$ phases measured by LASS are plotted in
Fig.~\ref{fig:solution0}.  There is a large offset in the
\swave, about 75$^{\circ}$, not seen in \pdash\ or \dwave s.
The shapes of \sdash\ and \pwave s are also not the same.
Unless significant admixture of $I=3/2$ $\Km\pip$ production
occurs, these results suggest that the conditions for the Watson
theorem are not met in these data.

\begin{figure}[hbt]
  \centerline{%
  \epsfig{file=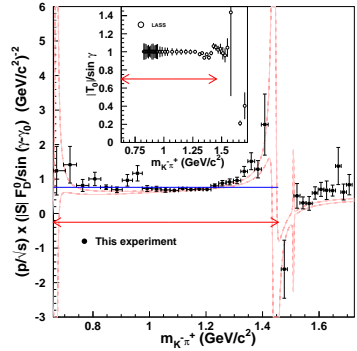,
               width=0.35\textwidth,angle=0}}
  \caption{The quantities 
   $p/\sqrt(\SAk)\times|\hbox{S}(\SAk)|\times
   F_{\sst D}(q,r_{\sst D},0)/\sin(\gamma_k-\gamma_0)$
   plotted as solid circles for each point obtained for the \swave\
   amplitude in the model-independent fit described in the text.
   Three points between 1400 and 1450~\MeVcc\ are omitted from
   the plot as their values for $\sin(\gamma_k-\gammaz)$ are
   very small, making their errors extremely large.
   The region between the dashed lines shows the one standard 
   deviation limits of this quantity for the \swave\ amplitude 
   obtained from the model-dependent fit.  The inset shows, as
   small open circles, the quantities
   $|T_0(\SA)|/\sin(\gamma)$ taken from the LASS experiment.  
   The elastic range, up to $K\eta^{\prime}$ threshold, is
   indicated by double-headed arrows in both plots.
  \label{fig:smag_pvec}}
 \end{figure}

The production function $Q(\SA)$ is shown in
Fig.~\ref{fig:smag_pvec}.  The scattering amplitude $T(\SA)$ is
assumed to be elastic, $T=\sin(\gamma-\gamma_0)$.  The phase
offset $\gamma_0$ is required to account for the difference
between elastic scattering and $D$ decay.  There appears to be 
significant \SA-dependence above $\sim~1.2~\GeVcc$ that grows
in the region near 1.4~\GeVcc\ where $\gamma\sim\gamma_0$.

\section*{Acknowledgments}
We thank the LASS collaboration for making 
their data available to us.
We gratefully acknowledge the assistance of the staffs of Fermilab
and of all the participating institutions.  This research was
supported by the Brazilian Conselho Nacional de 
Desenvolvimento Cient\'{\i}fico e Tecnol\'{o}gico,
CONACyT (Mexico), FAPEMIG
(Brazil), the Israeli Academy of Sciences and Humanities,
the U.S. Department of Energy, the U.S.-Israel
Binational Science Foundation, and the U.S. National Science 
Foundation.  Fermilab is operated by the Universities Research
Association for the U.S. Department of Energy.

 \bibliographystyle{aipproc}
 \bibliography{the_paper}

\begin{thebibliography}{8}
\expandafter\ifx\csname natexlab\endcsname\relax\def\natexlab#1{#1}\fi
\providecommand{\enquote}[1]{``#1''}
\expandafter\ifx\csname url\endcsname\relax
  \def\url#1{\texttt{#1}}\fi
\expandafter\ifx\csname urlprefix\endcsname\relax\def\urlprefix{URL }\fi
\providecommand{\eprint}[2][]{\url{#2}}

\bibitem[Aston et~al.(1988)]{Aston:1987ir}
D.~Aston, et~al., \emph{Nucl. Phys.} \textbf{B296}, 493 (1988).

\bibitem[Estabrooks et~al.(1978)]{Estabrooks:1977xe}
P.~Estabrooks, et~al., \emph{Nucl. Phys.} \textbf{B133}, 490 (1978).

\bibitem[Note]{conjugate}
Note, {Charge conjugate states are always implied unless explicitly
  stated otherwise.}

\bibitem[Aitala et~al.(2002)]{Aitala:2002kr}
E.~M. Aitala, et~al., \emph{Phys. Rev. Lett.} \textbf{89}, 121801 (2002),
  \eprint{hep-ex/0204018}.

\bibitem[Blatt and Weisskopf(1952)]{BlattWeiss}
J.~M. Blatt, and V.~F. Weisskopf, \emph{Wiley, New York} p. 361 (1952).

\bibitem[Tornqvist(1995)]{Tornqvist:1995kr}
N.~A. Tornqvist, \emph{Z. Phys.} \textbf{C68}, 647--660 (1995),
  \eprint{hep-ph/9504372}.

\bibitem[Aitala et~al.(2005)]{Aitala:2005yh}
E.~M. Aitala, et~al.  (2005), {Submitted to Phys. Rev. D.},
  \eprint{hep-ex/0507099}.

\bibitem[Watson(1952)]{Watson:1952ji}
K.~M. Watson, \emph{Phys. Rev.} \textbf{88}, 1163--1171 (1952).

\end{thebibliography}

\end{document}